\documentclass[%
 reprint,
 amsmath,amssymb,
 aps,
]{revtex4-1}

\usepackage{graphicx}
\usepackage{bm}
\usepackage{hyperref}

\begin{document}

\preprint{APS/123-QED}

\title{New method of soft modes investigation by Little-Parks effect}

\author{Victor D. Lakhno}
\email{lak@impb.ru}
\affiliation{%
Keldysh Institute of Applied Mathematics of RAS, 125047, Moscow, Russia
}%


\date{\today}

\begin{abstract}
As is known, Little-Parks effect concerned with oscillations of the critical temperature of a superconducting transition was one of the first effects which suggested the existence of Cooper pairing in conventional superconductors. It is shown that in high-temperature superconductors (HTSC) the Little-Parks effect based on bipolaron mechanism which in HTSC is the equivalent of Cooper pairing can be anomalously high. The results obtained can be used as a new method for detecting soft phonon modes in nanostructures on the basis of HTSC. The results can be also used to enhance the critical temperature of a superconducting transition.
\begin{description}
\item[PACS numbers]
74.25.Gz, 74.25.Kc
\end{description}
\end{abstract}

\pacs{Valid PACS appear here}
\maketitle




Little-Parks effect \cite{1} is related to the obvious fact that quantization of
a magnetic flux in a multiply connected superconductor (SC) is caused by relevant
quantization of the current around each cavity in SC. Quantization of the current
leads to oscillations of the velocity of superconducting current carriers.
Since the temperature of SC transition depends on the velocity of Bose condensate,
oscillations of its velocity lead to relevant oscillations of the temperature
of SC transition. Investigation of this phenomenon has become the focus
of attention in recent years with the prospect to use it for enhancing the critical temperature of SC \cite{2,3,4,5}.

Since at present a generally accepted theory of high-temperature superconductivity (HTSC)
 is lacking, a generally accepted treatment of the Little-Parks effect in these materials
is lacking either. The aim of this work is to develop a theory of Little-Parks effect
on the basis of translation invariant (TI) bipolaron theory of HTSC \cite{6,7}.

In TI-bipolaron theory of HTSC bipolarons are regarded as charged bosons placed in electron
Fermi liquid which screens an interaction between the bipolarons and the problem is reduced
to that of an ideal Bose-gas. Following \cite{6,7} let us consider an ideal Bose gas of TI
bipolarons which is a system of $N$ particles occuring in volume $V$. Let us write $N_0$ for
the number of particles in the lowest one-particle state and  $N'$ for the number of particles in higher states.

Let us proceed from the spectrum of excited states of a TI bipolaron which coincides with Landau's roton spectrum \cite{8}:

\begin{equation}
\label{eq:1}
\epsilon(\textbf{k})=\omega+k^2/2M
\end{equation}
where $\textbf{k}$ is the bipolaron wave vector, $\omega$ is the frequency of an optical phonon, $M=2m$, $m$
is the effective mass of a band electron, $\epsilon (0)=0$ corresponds to the ground state of
a TI bipolaron. According to \cite{9}, the excitation spectrum of Bose condensate
of TI bipolarons which moves in a magnetic field of intensity $\textbf{B}$ at the velocity  of $\textbf{u}$  will have the form:

\begin{eqnarray}
\label{eq:2}
\epsilon(\textbf{k})=\omega+\frac{k^2}{2M}+\frac{\eta}{M}\textbf{Bk}-\textbf{ku},
\end{eqnarray}
where $\eta$ is a certain constant. With the use of  \eqref{eq:2} we can write down an expression
for the number of bipolaron states $N$. On condition that $T<T_c$: $N=\sum_{k}m(\epsilon_k)$,
where $m(\epsilon_k)$  is a boson distribution function:
$m(\epsilon_k)=N_0\Delta(\textbf{k})+\left[\exp(\epsilon_k/T)-1\right]^{-1}\left(1-\Delta(\textbf{k})\right)$,
$\Delta (\textbf{k})=1$ for $\textbf{k}=0$ and $\Delta (\textbf{k})=0$  for $\textbf{k}\neq 0$ we get $N=N_0+N'$:

\begin{eqnarray}
\label{eq:3}
N'/V=(MT/2\pi\hbar^2)^{3/2}F_{3/2}(\tilde{\omega}/T),
\end{eqnarray}

\begin{eqnarray}
\label{eq:4}
F_{3/2}(\alpha)=\frac{2}{\sqrt{\pi}}\int_0^{\infty}\frac{x^{1/2}}{e^{x+\alpha}-1}dx,
\end{eqnarray}

\begin{eqnarray}
\label{eq:5}
\tilde{\omega}=\omega\left(1-\frac{u^2}{u^2_c}-\frac{B^2}{B_c^2}\right),
\end{eqnarray}

\begin{eqnarray}
\label{eq:6}
u_c=\sqrt{2\omega/M},\ \ B_c=\sqrt{2M\omega/\eta}.
\end{eqnarray}
where $\hbar$  is Plank constant. For $N'=N$, where $N$ is the total number of bipolarons,
equation \eqref{eq:3} determines the temperature of SC transition $T_c$. From general relations for quantization
of fluxoid $\Phi'$:  $\Phi'=n\Phi_0$, $\Phi_0=\pi\hbar c/e$, which in the case of bipolaron Bose condensate have the form:

\begin{eqnarray}
\label{eq:7}
\Phi'-\Phi=\frac{c}{2e}\oint M_{bp}\textbf{u}d\textbf{R},\ \ \Phi=\oint\textbf{A}d\textbf{R},
\end{eqnarray}
where $M_{bp}$ is the bipolaron mass, $e$ is the electron charge, $c$ is the lightspeed, $\Phi$
is the magnetic flux through the contour of integration $\textbf{R}$: $\Phi=\int\textbf{B}d\textbf{S}$.
In the case of a thin-walled cylinder corresponding to the Little-Parks experiment
we express the velocity of Bose condensate of TI bipolarons as:

\begin{eqnarray}
\label{eq:8}
u=\frac{\hbar}{M_{bp}R}\left(n-\frac{\Phi}{\Phi_0}\right),
\end{eqnarray}
	where $R$ is the cylinder radius. Equation \eqref{eq:8} describes oscillations of the velocity
of Bose-condensate of TI bipolarons and leads to oscillations of the superconducting transition  temperature \cite{10}.
Substituting \eqref{eq:5}, \eqref{eq:8} into the equation for the SC transition temperature \eqref{eq:3} and assuming
$\tilde{\omega}=\tilde{\omega}_0+\Delta\tilde{\omega}$,
where $\tilde{\omega}_0=\omega(1-\Phi^2/\Phi^2_c)$, $\Delta\tilde{\omega}=-\omega u^2/u^2_c$,
and $\Phi_c$ is the critical magnetic flux corresponding to the critical field $B_c$,
for small deviation of $\Delta\tilde{\omega}$ we express the critical temperature deviation from \eqref{eq:3} as:
 \begin{eqnarray}
\label{eq:9}
\frac{\Delta T_c}{T_c}=\frac{\xi ^2}{R^2}\left(n-\frac{\Phi}{\Phi_0}\right)^2,
\end{eqnarray}

\begin{eqnarray}
\label{eq:10}
\xi=-\frac{1}{3(2\pi)^{3/2}}\frac{M^2}{M^2_{bp}}\frac{1}{n^{2/3}_{bp}}
\left(\frac{MT_c}{n^{2/3}_{bp}\hbar^2}\right)^{1/2}\textit{Li}_{1/2}(e^{-\alpha}),
\end{eqnarray}
	
\begin{eqnarray}
\label{eq:11}
\textit{Li}_{1/2}(e^{-\alpha})=\frac{1}{\sqrt{\pi}}\int^{\infty}_0\frac{1}{\sqrt{t}}\frac{dt}{e^{t+\alpha}-1},
\end{eqnarray}

\begin{eqnarray}
\label{eq:12}
\alpha=\frac{\omega}{T_c}\left(1-\frac{\Phi^2}{\Phi^2_c}\right),
\end{eqnarray}
where $n_{bp}$ is the concentration of TI bipolarons. Formula \eqref{eq:9} determines the amplitude
of oscillations $T_c$  which take place as the magnetic flux $\Phi$ changes. Notwithstanding
the oscillating behavior of $\Delta T_c$ on magnetic flux, expression \eqref{eq:5} suggests that there exists
a maximum possible value of the magnetic flux $\Phi_{max}=\Phi_c=\pi R^2B_c$  corresponding to the maximum possible
magnetic field for which a superconductivity of a Bose condensate is possible with
$T_c(\Phi=\Phi_c,u=0)=3,31\hbar^2n^{2/3}_{bp}/M$.

At "low" critical temperatures: $\alpha>>1$, $\textit{Li}_{1/2}(e^{-\alpha})\cong(e^{-\alpha})$,
while at high ones $\alpha<<1$, $\textit{Li}_{1/2}(e^{-\alpha})\approx {\sqrt{\pi/\alpha}}$.
With the use of expressions \eqref{eq:9}, \eqref{eq:10}  for $\alpha\cong1$, $n_{bp}=10^{19}$cm$^{-3}$ \cite{6}, $m=m_0$, $M\approx M_{bp}$,  $R=100$nm,
we will estimate $\Delta T_c/T_c$  as:  $\Delta T_c/T_c\approx 10^{-4}$,
which is close to the estimate resulting from the BCS theory \cite{10}. It also follows from expressions \eqref{eq:9}, \eqref{eq:10} that for
$\omega /T_c<<1$ (accordingly $\alpha<<1$) the ratio $\Delta T_c/T_c\approx 10^{-4}(T_c/\omega)^{1/2}$
can be much greater than that obtained on the basis of the BCS theory.

The Little-Parks effect is usually applied to verify the fundamentals of the superconductivity theory, in particular, to prove the phenomenon of pairing of current carriers in superconductors. The results obtained suggest that the Little-Parks effect can be also used to determine the presence of soft phonon modes for which  $\omega\rightarrow 0$ \cite{11} which are associated with structural instability and structural phase transitions in HTSC. In this case the oscillations of $\Delta T_c/T_c$ can be abnormally high.
 	
Presently, the main methods for investigation of phonon spectra are inelastic x-ray scattering  and
inelastic neutron scattering \cite{12,13}. The accuracy of these methods is limited to several meV.
As a result, the estimated value of $(T_c/\omega)^{1/2}$ will yield a small coefficient:
$<10$ - for real values of $T_c$ equal to several dozens of degrees and may be in contradiction
with the data on $\Delta T_c/T_c$, obtained on the basis of the Little-Parks approach.

This conclusion is supported by abnormal softening of phonon modes in the family of
$\textit{La}_{2-x}\textit{Sr}_x\textit{CuO}_4$  compounds \cite{11} with values of x close
to those for which the ratio ${\Delta T_c}/{T_c}$ was abnormally high \cite{2} which were associated in \cite{2}
with the possibility of the formation of vortices and their interaction with the oscillation current.

The results obtained are based on the idea of Bose condensation of TI bipolarons possessing
a small correlation length $(\approx 1 nm)$. This is valid only for the case of strong electron-phonon
interaction (EPI) when the EPI constant $\alpha _{ph}$ is large. The condition $\alpha _{ph}\approx \infty$
is fulfilled only for  $\omega\rightarrow0$, since  $\alpha _{ph}\approx\omega^{-1/2}$.
Hence, the bipolarons formed by such modes will have the lowest energy as compared to other phonon
branches and, therefore, just such phonons will form the Bose condensate considered which leads
to anomalies in the Little-Parks effect.

Expressions \eqref{eq:9}--\eqref{eq:12} also suggest that generally when the phonon mode softening is not large,
in an effort to achieve the maximum amplitude of $T_c$ oscillations in HTSC materials one should
use the maximum value of the magnetic flux $\Phi\rightarrow\Phi_c$. In this case an arbitrarily
small deviation of the Bose condensate velocity from its equilibrium value $u=0$ leads to a finite change in $T_c$.



\end{document}